# Physics Graduate Student Employment: What We Can Learn From Professional Social Media

Erin M. Tonita[1], Ludmila Szulakowska[2], Joshua Baxter[1], Erin L. Flannigan[1], Naveena Janakiraman Narayanan[3], and Jean-Michel Ménard[1]

[1]Department of Physics, University of Ottawa, Ottawa, Canada
[2]Stewart Blusson Quantum Matter Institute, University of British Columbia, Vancouver, Canada
[3]Department of Chemical and Biological Engineering, University of Ottawa, Ottawa, Canada

## I. Introduction

In science, and especially in the physics community, motivation for pursuing post-secondary education can be quite diverse. Whether our inspiration to study physics came from curiosity about the universe around us, a love of applied math, or a desire to pursue a specific career, we are all eager to see how far our education will actually take us when we enter the job market. Will we find work in a physics-related discipline or use our newly acquired skills to contribute to a different sector? The last decade has seen major changes in our daily lives and the job industry, and many students who are about to start their post-secondary education, or to enroll into a graduate program, may wonder how well recent physics graduates have performed in this evolving job market. More specifically, what are recent physics Master's (MSc) and Doctoral (PhD) graduates doing now? Also, considering the increasing influence of social media in our lives, can we assess how the size of a virtual network may have an impact on employment?

In this study, we have addressed these questions by looking at the professional path of recent physics graduate students at the University of Ottawa (uOttawa), our home university. Our database was populated from the public online repository of MSc and PhD theses submitted between the academic years of 2011 to 2019, with employment information subsequently collected from the professional social media platform LinkedIn. Due to the popularity of social media among recent graduates, we could collect detailed professional information from 80% of the 113 graduates considered in this study. In comparison, previous studies on this subject mostly relied on cold calling for data collection and reported a response rate around 50% [1]. We discuss our data collection procedure and analysis in the Methodology section following this article.

## II. Results and Discussion

Before analyzing employment details, we first looked at the graduates' academic histories. Our goal was to determine the percentage of MSc students that chose to continue to the PhD level, and among them, how many decided to move to a different university. Figure 1a shows that more than a half of MSc students decided to pursue a PhD. Slightly fewer than half of these students (23%) moved to a different university after completing their MSc degree. This level of student mobility after the MSc is comparable to the one observed in US universities [2]. The fact that many students stay at the same university after their MSc is reflected in our data by the relatively large ratio (41%) of PhD graduates at uOttawa who also completed a MSc degree at uOttawa (Fig. 1b). Note that 37% of the PhD students did not provide MSc information, indicating that many of them may have fast-tracked to their PhD (1-year MSc without a thesis before entering their PhD program) or made a direct entry after their undergraduate degree. Finally, the average time for completing a MSc degree was 2.4 years and 5.0 years for a PhD degree. These results included all graduates, both full-time and part-time students, as well as those who took time off for reasons such as parental leave. We found that 60% of uOttawa MSc students and 70% of the PhD candidates graduate within 2 and 5 years, respectively, which are standard time frames for Canadian universities.

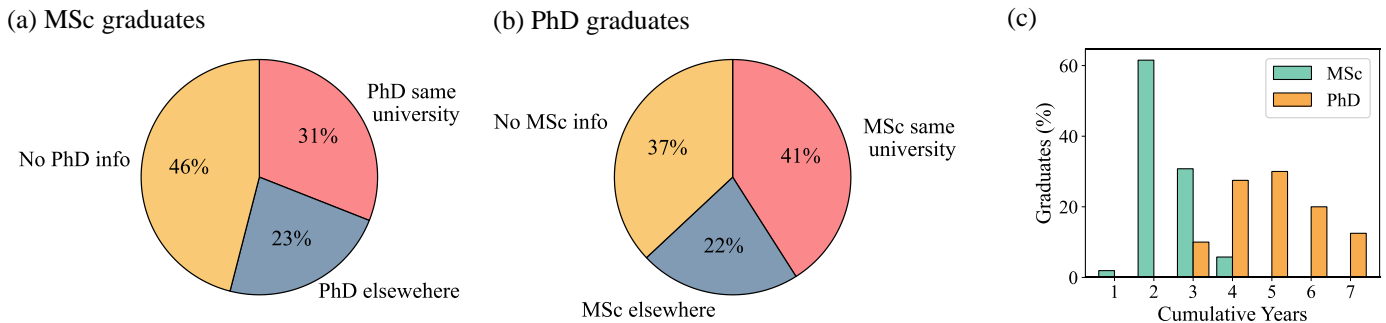

Fig. 1. Graduate school history showing (a) the ratio of MSc graduates from uOttawa pursuing a PhD degree at the same university, or elsewhere, and (b) the ratio of PhD graduates at uOttawa who previously obtained a MSc degree from uOttawa (same university), or elsewhere. (c) Length of MSc and PhD degrees at uOttawa indicated by the number of students who graduated within a certain time period.

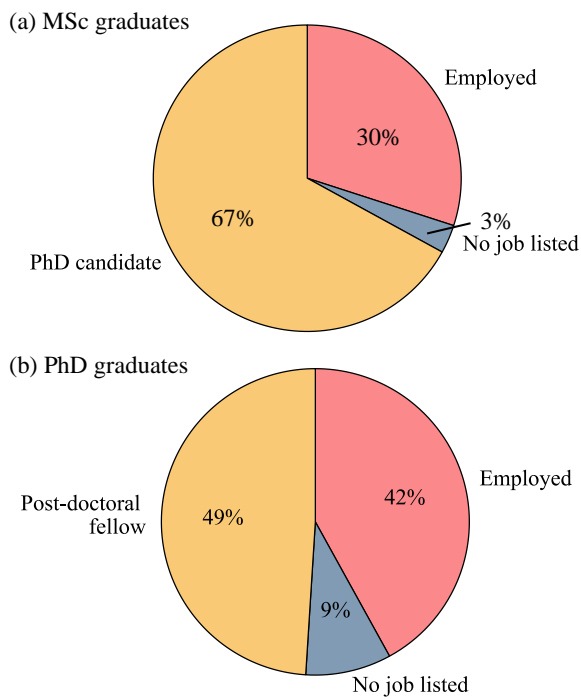

Fig. 2. LinkedIn employment status one year following graduation for a) MSc graduates and b) PhD graduates.

Detailed career information on 89 graduates was obtained from public information on LinkedIn. We focused on the job status within one year following graduation because it provides a strong indication of the degree of employability immediately after receiving the diploma. Note that we are unable to account for graduates who willingly decided to take time off, possibly for traveling, parental leave or other personal reasons. As illustrated in Fig. 2a, most MSc graduates continued to the PhD level, 30% of the MSc graduates were employed and only 3% had no job listed after one year of graduation. A lack of employment information does not necessarily indicate unemployment. Here, we simply refer to graduates falling into this category as having 'no job listed'. This designation also includes anyone pursuing further professional specializations, such as medical school. Figure 2b shows that 49% of the PhD graduates found a post-doctoral research position within a year, 42% occupied another job, while 9% had no job listed after one year of graduation. This information is promising for current physics graduates as it shows that 94% of physics graduates are employed or continuing their post-graduate education after one year of receiving a graduate degree. Of the remaining 6%, approximately a third are actively pursuing a subsequent professional degree, such as medical or law school. This ratio is consistent with a recent study on PhD graduates from the United States [1]. One could argue that graduates are more likely to share their success stories on social media, which would imply that our ratio of graduates with "no job listed" is slightly underestimated. However, it is also possible that those without a job are more likely to use social media to establish a network and to find a job, which would indicate that our ratio of graduates with "no job listed" might be, on the contrary, overestimated.

The 2019 employment status of recent MSc and PhD physics graduates was also obtained and categorized into career sectors. We chose six job categories corresponding to academia, industry, and government research, which are traditional science-related fields, as well as information technologies (IT), teaching and 'other'. The 'teaching' category encompasses all those in teaching professions outside of university institutes, such as in high schools or colleges, with university positions considered in the 'academia' category. Of those in the academia category, approximately half are in post-doctoral fellowships, while the remainder are employed in research positions at an academic institution. Our results are presented in Fig. 3. Unsurprisingly, academic and industry careers account for most physics-graduate jobs, with ~53% of recent graduates having employment in these two areas. Notably, 40% of graduates end up in an academic or teaching environment, in agreement with a recent report [3]. The third largest career sector is 'other'; 20% of physics graduates fall into this diversified category that includes jobs in patent agencies, health and medicine, consulting, and business management. Similar to other studies conducted on physics graduates [4], our results show that uOttawa physics graduates are able to diversify their experience and transfer their skills to broader job markets.

In addition to the current employment of recent graduates, we investigated the general online networking strategies adopted by physics graduates to conquer the job market. LinkedIn provides a quantitative measure of this activity through the number of connections. A single LinkedIn 'connection' corresponds to an online contact that has been made between the graduate and another person for the purpose of sharing employment opportunities and skill sets. We analyzed these numbers for uOttawa graduates and looked for correlations between connectivity and career choices. Figure 4 shows the number of LinkedIn connections of uOttawa graduates while also indicating their 2019 employment information. Most graduates

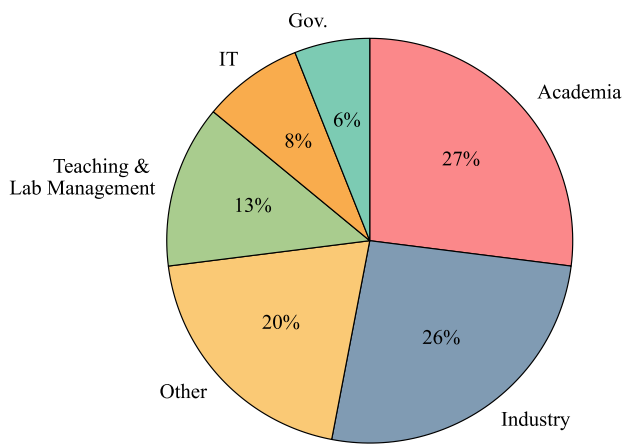

Fig. 3. Employment sectors for physics graduates.

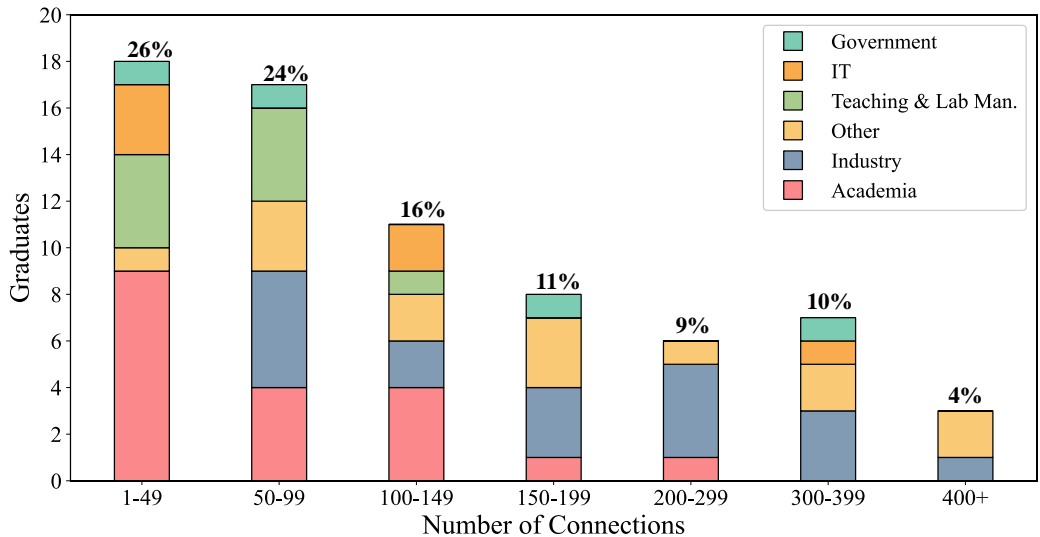

Fig. 4. Correlation between LinkedIn connections and current employment career sector.

have fewer than 100 connections, and only 4% of graduates have more than 400 connections.

Though not a perfect predictor, the number of LinkedIn connections can be an indicator of how often one uses the site, either passively (consuming content) or actively (producing/sharing content). It has been shown that use-frequency is a good predictor of how much one will benefit from LinkedIn, with the greatest benefit occurring when all users in a network know the interests and capabilities of one another [5].

Employment information shows that graduates working in academia (red) tend to have fewer online connections in comparison to those in industry (blue) and the 'other' category (yellow). This suggests that, at the time of our study, in-person interactions and networking were potentially more important to finding jobs in academia. In industry, we may assume that a large professional network is more useful to finding a job, or more valued by employers. For graduates with a career in the 'other' category, connection numbers reach the highest, potentially indicating a need to diversify their networks beyond science and engineering. As results in Fig. 4 show that workers in 'industry' and 'other' categories have the largest number of connections on LinkedIn, these people may be more likely to have a professional social media profile. Therefore, our findings could be slightly more weighted towards graduates who are employed in these career sectors.

## III. CONCLUSION

In this study, we investigated the employment status of recent uOttawa physics MSc and PhD graduates, finding that 94% of graduates from 2011-2019 are either employed or pursuing further physics education one year post-graduation. Our results, in agreement with previous studies on physics graduates, highlight that graduates primarily find employment quickly and in their field of study, with most graduates employed in either academia or physics-related industries. We also found that a significant portion of employed graduates, 20%, find employment in non-traditional physics careers, such as business management and healthcare. This may be because, as previous studies have found, physics graduates are well prepared to pursue many careers and are sought for their flexibility, problem-solving skills, and exposure to a range of technologies [6]-[8]. In addition, we explored the role of social media in employment, and argued that online networking via social media platforms is a potential method of creating valuable professional connections. Graduates with careers in academia tend to have lower online connectivity compared to graduates with careers in industry or non-traditional fields, suggesting a greater importance for online networking for students interested in non-academic careers.

We would like to conclude this article on a personal note. The authors of this article are, for the most part, graduate students who grew up being inspired by popular science and the mainstream interpretation of what physicists study: astrophysics, cosmology, string theory, etc. After years of study, many physicists find that their passions lie in less popularized fields. We are researchers studying condensed matter physics, photonics, nuclear physics, and biophysics, and we truly love what we do. Though it is not always obvious to those commencing on their academic journey what careers are available to them when they finish, our study has revealed that there are numerous opportunities in both traditional and non-traditional physics paths. While our initial goal was to inform current and future uOttawa graduates about recent graduate employment, we think these results are of interest to the entire physics community. We suggest that this exercise is routinely performed in every department and university to maintain a record of graduate employment and provide relevant and up-to-date information on career prospects for future students.

## IV. METHODOLOGY

We primarily collected data for this study from LinkedIn, a professional online networking platform that fosters connections between people for the purpose of sharing employment opportunities, as well as individuals' employment statuses and skills. When two people 'connect' on LinkedIn, they can view each other's resumes and skill sets, and pass along employment opportunities. Note that information on LinkedIn is likely to accurately reflect the curriculum vitae presented to potential employers, therefore we believe that our results are highly reliable and arguably more trustworthy than those that can be obtained from online surveys or phone surveys. Data was supplemented with information from ResearchGate, a similar platform for research-oriented networking. We limit our assessment of available career information to employment status and career-sector categorization, rather than attempt to assess more subjective career information like job quality.

Out of the 136 uOttawa physics theses submitted by 113 individuals between January 2011 and February 2019, 78 were MSc and 59 were PhD theses. We were able to obtain LinkedIn education information on 103 graduates and employment information on 89 graduates. Note that all percentiles presented in this article have been rounded to whole numbers. Finally, the total number of graduates considered in this study contains 27% women and 73% men as estimated from Gendre-API.com, an internet platform that predicts gender from the full name. This method cannot account for non-binary students, transgender students enrolled under their birth name or students with names that can apply to both men and women. As such, it provides only an incomplete overview on the gender ratio. A study relying on self-identification would be necessary to obtain more reliable data.

ACKNOWLEDGEMENT OF PUBLICATION

This manuscript has been accepted for publication in *Physics in Canada*.